\documentclass[letterpaper, 10 pt, conference]{ieeeconf}
\IEEEoverridecommandlockouts 
\usepackage{tikz}
\usepackage{dsfont}

\usepackage{amsthm}
\usepackage{amssymb}
\usepackage{romannum}
\usepackage{xcolor}
\usepackage{cite}
\usepackage{amsmath,amssymb,amsfonts}
\usepackage{textcomp}
\usepackage{graphicx} 
\usepackage{algorithm}
\usepackage[capitalize]{cleveref}
\usepackage{algpseudocode}
\usepackage{bbm}

\renewcommand{\baselinestretch}{0.93}

\usepackage{pgfplots}
\usepackage{pifont}
\pgfplotsset{compat=1.8}
\usepackage{textcomp}
\theoremstyle{definition}

\newtheorem*{assumption*}{Assumption}
\newtheorem{theorem}{Theorem}

\newtheorem{lemma}[theorem]{Lemma}

\newtheorem{definition}{Definition}
\theoremstyle{remark}
\newtheorem*{remark}{Remark}

\usepackage{bbding}

\usetikzlibrary{patterns}
\usepackage{pgfplots}
\pgfplotsset{compat=1.8}
\usepgfplotslibrary{statistics}
\usepackage{url}
\newcommand{\tup}[1]{\left(#1\right)}
\def\BibTeX{{\rm B\kern-.05em{\sc i\kern-.025em b}\kern-.08em
    T\kern-.1667em\lower.7ex\hbox{E}\kern-.125emX}}
\begin{document}
\title{Co-investment with Payoff Sharing \\Benefit Operators and Users in Network Design}
\author{Mingjia He$^{1}$, Andrea Censi$^{1}$, Emilio Frazzoli$^{1}$, and Gioele Zardini$^{2}$
\thanks{This work was supported by the ETH Zürich Mobility Initiative (MI-03-22) and ETH Zürich Foundation project number (2022-HS-213).}
\thanks{$^{1}$Institute for Dynamic Systems and Control, ETH Zürich, 8092 Zürich, ZH, Switzerland 
(e-mail: minghe@ethz.ch; acensi@ethz.ch; emilio.frazzoli@idsc.mavt.ethz.ch). }
\thanks{$^{2}$Laboratory for Information and Decision Systems, Massachusetts Institute of Technology, Cambridge, MA, USA. 
(e-mail: gzardini@mit.edu).}
}

\maketitle

\begin{abstract}
Network-based complex systems are inherently interconnected, with the design and performance of subnetworks being interdependent. 
However, the decisions of self-interested operators may lead to suboptimal outcomes for users.
In this paper, we consider the question of what cooperative mechanisms can benefit both operators and users simultaneously.  
We address this question in a game theoretical setting, integrating both non-cooperative and cooperative game theory. 
During the non-cooperative stage, subnetwork decision-makers strategically design their local networks. 
In the cooperative stage, the co-investment mechanism and the payoff-sharing mechanism are developed to enlarge collective benefits and fairly distribute them. 
A case study of the Sioux Falls network is conducted to demonstrate the efficiency of the proposed framework. The impact of this interactive network design on environmental sustainability, social welfare and economic efficiency is evaluated, along with an examination of scenarios involving regions with heterogeneous characteristics.
\end{abstract}


\section{Introduction}
\label{sec:introduction}
Globalization has strengthened the interconnections of economic, political, and technological complex systems. 
Networks form the backbone of such complex systems, facilitating the movement of people, goods, and information.
According to the World Bank, global trade as a percentage of gross domestic product (GDP) increased from 50\% in 2000 to 63\% in 2022 \cite{globaltrade}. 
The growing interconnections, coupled with the expanding world population as of 2024, have a transformative impact on real-world systems including healthcare, transportation, communication, production, and energy.
In many of such complex systems, networks are typically composed of multiple subnetworks, each characterized by distinct decision-makers.
Due to the network inter-connectivity, the design of one subnetwork may affect the performance of others, and vice versa.  
In this context, it is essential to account for the \emph{strategic interactions} among subnetwork designers.

As self-interested agents may lead to system suboptimality, it is essential to foster cooperation in the multi-agent environment.
The key questions include: when should cooperation occur? What agreements on network design should be reached? What are the individual gains that should be derived from such cooperation?
Furthermore, for real-world systems, the decision-making process often presents both competition and cooperation. 
For instance, in the supply chain, competing suppliers may collaborate by sharing manufacturing or transportation capacity to improve cost efficiency~\cite{simatupang2002collaborative}. 
In economics, governments manage their domestic economies while simultaneously negotiating trading agreements with other nations and international institutions~\cite{irwin2024does}.
Similarly, for the network design problem, subnetwork designers can make independent decisions regarding their own capacities, and, meanwhile, they may negotiate and seek joint design plans to achieve mutual benefits.

\begin{figure}[tb]
    \centering\includegraphics[width=1\linewidth]{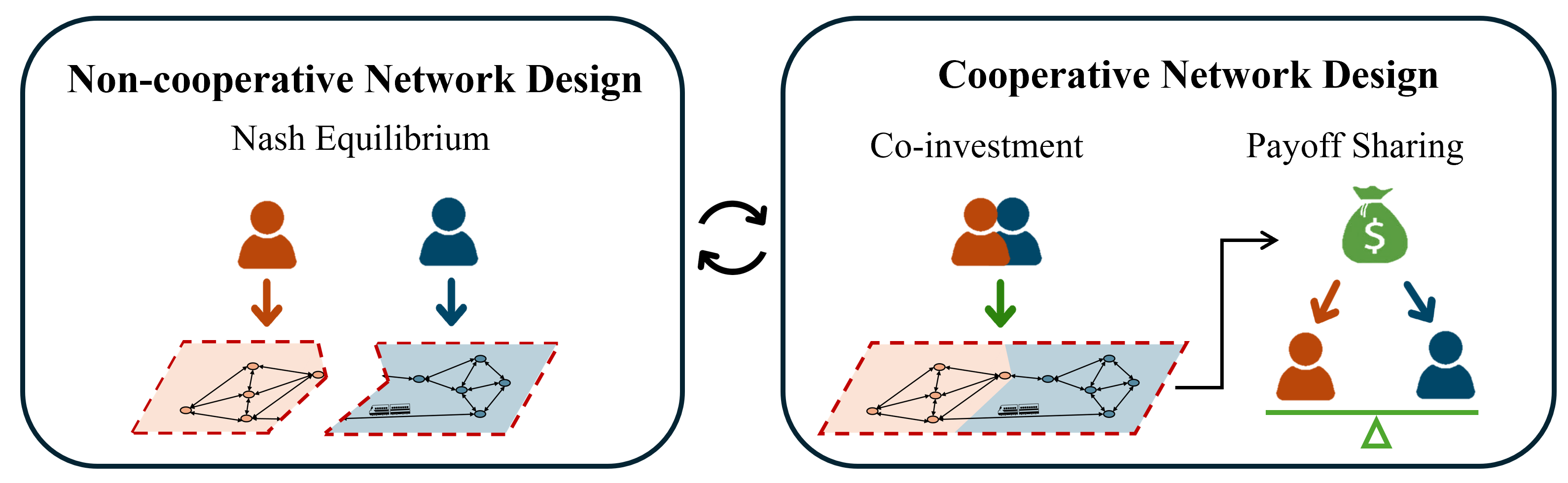}
    \caption{The interactive network design framework, featuring a non-cooperative, as well as a cooperative phase.}
    \label{fig:co_framework}
\end{figure}

Game theory is a powerful tool that has been utilized to model strategic interactions in various multi-agent settings (see, e.g.,~\cite{zardini2021game,zardini2023strategic}~for previous work in our lab). 
Specifically, one can typically classify games into \emph{cooperative} or \emph{non-cooperative}. In the non-cooperative design, agents make decisions independently, optimizing their individual outcomes without communication. 
In cooperative games, instead, agents can choose to engage in joint efforts to enhance their collective outcomes in the competitive environment.
Considering the research problems, we propose a game theoretical framework for interactive network design (see \cref{fig:co_framework}). This framework accounts for the characteristics of both cooperative and non-cooperative games.
\emph{Co-investment} and \emph{payoff-sharing} mechanisms are designed to search for optimal agreements and to distribute collective benefits, which can benefit both the subnetwork operators and system users.
\subsection{Related research}
In transportation planning, network design is a crucial strategic decision which significantly impacts the operation of mobility systems~\cite{FARAHANI2013281,liu2021review,zardini2022co,zardini2022analysis}. The network design problem typically involves modifying an existing network by adding new connections and expanding the capacity of existing ones. 

Strategic interactions in network design problems can be categorized into two types: network-user, and network-network ones. 
Network-user interactions lie at the interface between network designers and network users. 
Specifically, network operators can adopt user-oriented objectives such as minimizing total travel time~\cite{gao2005solution}, maximizing the number of customers served~\cite{lo2009time} and profit~\cite{dimitriou2008genetic}. 
In turn, travelers make their decisions on transport modes and routes based on the available network infrastructure.
Network-network interactions refer to the interdependence among the decisions of subnetwork designers. In a transportation network composed of multiple subnetworks, each can be managed by different operators (e.g., administrative authorities). Each authority may have distinct design goals and varying levels of funding for their respective subnetwork. 
Such reasoning applies to various geographical scales, for instance, international networks that connect multiple countries~\cite{itf2023,liu2023global}, and urban-rural networks that involve different municipalities~\cite{porru2020smart}. 

Extensive past research models the interaction between network design and users’ decisions.
Specifically, the network design problem has been often modeled as a bi-level problem. 
It involves an upper-level problem where the operator first plans the transport network, and a lower-level problem for travelers’ decisions based on the designed infrastructure.
The most prevalent low-level models include traffic assignment models, discrete choice models, and combinatorial models integrating the features of both.
User Equilibrium (UE) traffic assignment assumes that users make decisions individually.
In a Social Optimum (SO) traffic assignment, travelers cooperate to minimize the total system travel time. 
In deploying autonomy-dedicated facilities, Zhang et al.~\cite{ZHANG2023102784} used a stochastic UE model for the route choice behavior in a mixed traffic environment.

Moreover, the range of traveler choices is broadened in the multi-modal transportation network design problem. 
Chow and Sayarshad developed a multi-objective framework to model the symbiotic relationships between coexisting networks~\cite{chow2014symbiotic}. 
This framework is applied to the design of bike sharing networks in downtown Toronto.
Wang considered the integration of the road, transit, and bike subnetworks, and proposed a tri-level network design problem to maximize network capacity while accounting for mode split and traffic assignment~\cite{wang2022large}. 
Authors in~\cite{wang2023integrated} designed a multi-modal network that includes rail, truck, and maritime services. This network is designed to accommodate shippers with varying demands and service preferences, with the aim of minimizing the total cost of shipping.

Exploring network-user interactions has provided valuable insight into single-region network design problems. 
However, the understanding of network-network interactions remains insufficient. 
\cite{medeiros2019cross,cb2022} concluded that cross-border railway services tend to be overlooked in investments of countries, leading to a potential loss of travel demand and undermining the competitiveness of rail transport. 
Moreover, a centralized network design approach neglects potential interests misalignment for local authorities, and the proposed strategies may result in being unrealistic to implement.
If each subnetwork focuses solely on its own design, it may result in suboptimal outcomes for the overall system due to the existing interconnections. 
Therefore, it is crucial to model interactions and facilitate cooperation in multi-regional transport network design.
This paper aims to establish a game-theoretic framework for the interactive network design problem and address questions of what cooperative mechanisms to consider for investment and payoff division. 
\subsection{Statement of Contribution}
The contribution of this work is threefold. 
First, we propose an interactive network design framework, taking into account both the cooperation and non-cooperation of operators. 
Second, we develop cooperative mechanisms of co-investment and payoff-sharing to explore cooperative opportunities to benefit both operators and users. 
Third, we conduct a case study on the Sioux Falls network and demonstrate the efficiency of the proposed framework. This work offers insights into the strategic interactions in the multi-regional network design and informs decision-makers on when and how to cooperate while respecting their specific regional network goals.

The paper is organized as follows. In Section \ref{sec:method} we introduce the key concepts in the interactive network design problem and develop an interactive network design framework in Section \ref{sec:coop}. Numerical experiments are conducted in \ref{sec:exp} and conclusions are drawn in \ref{sec:conclusion}.

\section{Method} \label{sec:method}
In this section, we define the key components and propose an
\emph{interactive network design} framework with co-investment and payoff-sharing mechanisms.

\subsection{Modeling the Mobility Network and Travel Demand}
\begin{figure}[tb]
    \centering\includegraphics[width=0.9\linewidth]{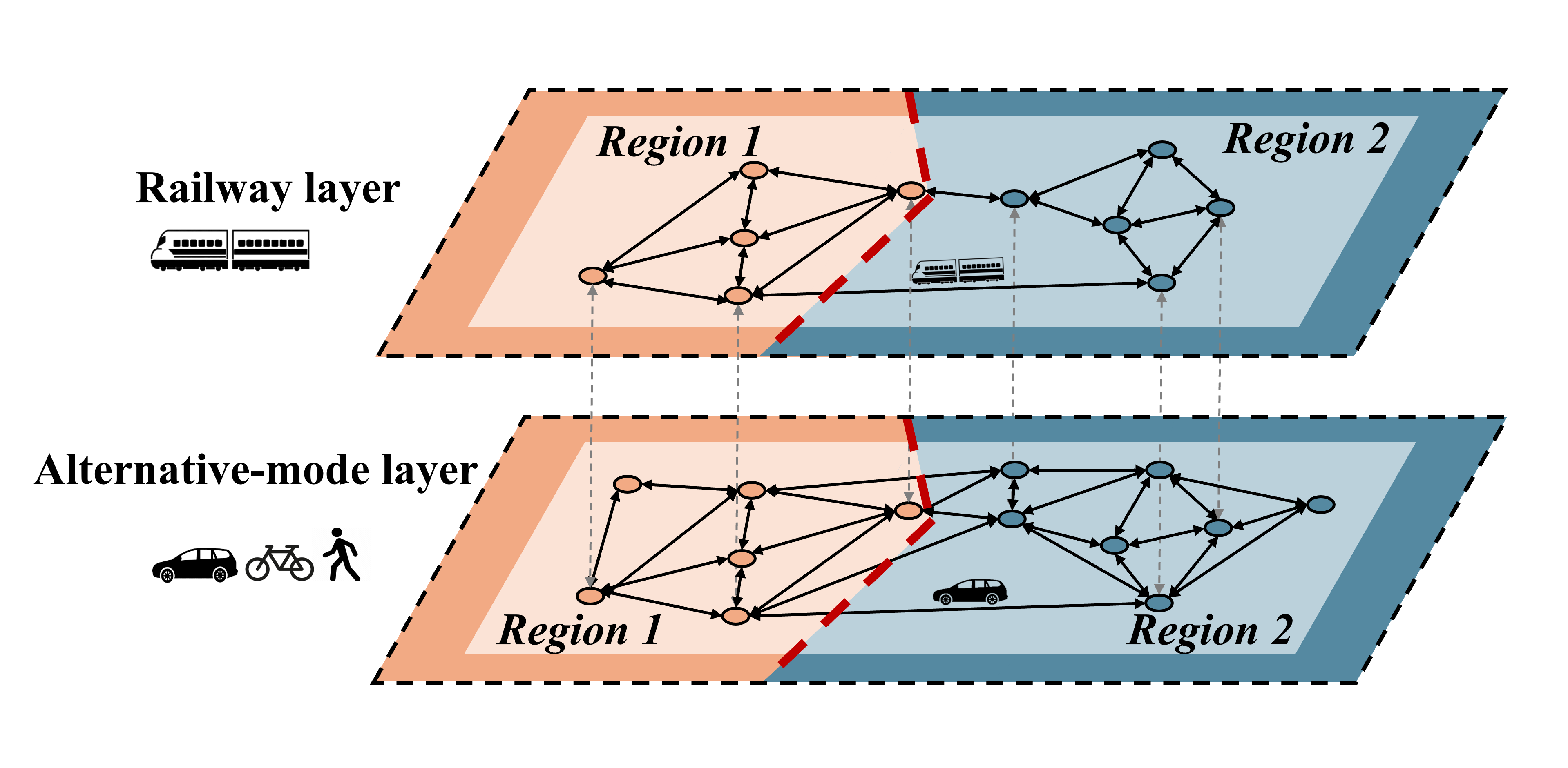}
    \caption{The interconnected multimodal mobility network for Region 1 and Region 2.}
    \label{fig:graph}
\end{figure}

We model the mobility network as an edge-labeled directed graph~$\mathcal{G}= (\mathcal{V}, \mathcal{E}, \mathcal{L})$, where~$\mathcal{V}$ is the set of vertices,~$\mathcal{E} \subseteq \mathcal{V} \times \mathcal{V}$ is the set of directed edges and~$\mathcal{L}: \mathcal{E} \to  \mathcal{Z} $ is a mapping from the set of edges $ \mathcal{E}$ to the set of edge labels $\mathcal{Z}$.
For a graphical example, see \cref{fig:graph}.
Specifically, an element~$z_e=\tup{x_e,c_e,l_e,t_e}\in \mathcal{Z}=\{0,1\}\times \left(\mathbb{N}_0 \cup \{\infty\}\right)\times \mathbb{R}^+\times \mathbb{R}^+$ is characterized by the availability of the mobility service on edge $x_e$, the capacity on the edge $c_e$, the edge length $l_e$, and the travel time associated to the edge~$t_e$.

In this work, we consider two authorities.
The graph \(\mathcal{G}\) can be divided into two subgraphs \(\mathcal{G}^1 = (\mathcal{V}^1, \mathcal{E}^1, \mathcal{L}^1)\) and \(\mathcal{G}^2 = (\mathcal{V}^2, \mathcal{E}^2, \mathcal{L}^2)\) corresponding to two regions (denoted Region 1 and Region 2), where  $\mathcal{V}^1$ and $\mathcal{V}^2$ are disjoint subsets of $\mathcal{V}$ satisfying $\mathcal{V}= \mathcal{V}^1 \cup \mathcal{V}^2$ and $\mathcal{V}^1 \cap  \mathcal{V}^2 = \emptyset$. The sets of edges for the subgraphs are defined as follows. The edge set of Region $i$ is $\mathcal{E}^i=\{(u,v) \in \mathcal{E}|u,v \in \mathcal{V}^i\}$ for $i \in I= \{1,2\}$, and the region-connecting edge set is defined as $\mathcal{E}^c=\{(u,v) \in \mathcal{E}| u \in \mathcal{V}^i, v \in \mathcal{V}^j, i,j \in I, i \neq j \}$. 
The edge sets satisfy $\mathcal{E}= \mathcal{E}^1 \cup  \mathcal{E}^2  \cup \mathcal{E}^c $ and $\mathcal{E}^1 \cap  \mathcal{E}^2 \cap \mathcal{E}^c = \emptyset$. 
This partition allows each regional network to be designed by regional authorities while maintaining the overall connectivity of the mobility network.
To enable multimodal mobility choices, each regional subgraph $(\mathcal{G}^i)_{i \in \{1,2\}}$ contains a railway network layer $\mathcal{G}^i_{R}= (\mathcal{V}^i_{R}, \mathcal{E}^i_{R}, \mathcal{L}^i_{R})$ and an alternative-mode network layer $\mathcal{G}^i_{A}= (\mathcal{V}^i_{A}, \mathcal{E}^i_{A}, \mathcal{L}^i_{A})$, which we assume represents an aggregated layer for other transportation modes such as private vehicles, bikes and walking, where $\mathcal{V}^i_{R} \cap \mathcal{V}^i_{A} = \emptyset$. 
The railway networks are characterized by stations $u \in \mathcal{V}^i_{R}$ and line segments $\tup{u,v} \in \mathcal{E}^i_{R}$; the network for the alternative-mode layer is modeled by intersections $u \in \mathcal{V}^i_{A}$ and link segments $\tup{u,v} \in \mathcal{E}^i_{A}$.
The mode-transfer edges set are represented as $\mathcal{E}^i_{C} \subseteq \mathcal{V}^i_{R} \times \mathcal{V}^i_{A} \cup \mathcal{V}^i_{A} \times \mathcal{V}^i_{R}$, allowing the switch of transportation mode during a single trip. 
Similarly, the region-crossing edge set consists of three subsets of edges: train edges, alternative-mode edges and mode-transfer edges, i.e., $ \mathcal{E}^c = \mathcal{E}^c_{A} \cup \mathcal{E}^c_{R} \cup \mathcal{E}^c_{C}$.

Given the above definitions, it holds that  $\mathcal{V} =  \mathcal{V}^1_{R} \cup    \mathcal{V}^1_{A} \cup  \mathcal{V}^2_{R} \cup  \mathcal{V}^2_{A} $ and $ \mathcal{E} =  \mathcal{E}^1_{R} \cup   \mathcal{E}^1_{A}\cup   \mathcal{E}^1_{C} \cup  \mathcal{E}^2_{R} \cup   \mathcal{E}^2_{A} \cup   \mathcal{E}^2_{C}\cup  \mathcal{E}^c_{R} \cup   \mathcal{E}^c_{A} \cup   \mathcal{E}^c_{C}$. By defining distinct subgraphs for regions and layers for railway and alternative modes, the overall structure can effectively support the modeling of a multiregion, multimodal transportation network. Local authorities can manage their respective regional networks, while customers can travel between regions and utilize the various modes of transportation available. 
A travel request is defined as $r_m=(o_m,d_m,\alpha_m,\theta_m) \in \mathcal{R} = \mathcal{V}_{A} \times \mathcal{V}_{A} \times \mathbb{N}_0 \times \Theta$, characterized by the origin $o_m$, the destination $d_m $, the number of trips $\alpha_m$ and the type of trips $\theta_m \in \Theta := \{\Theta^1_\mathrm{intra}, \Theta^2_\mathrm{intra},\Theta^1_\mathrm{inter}, \Theta^2_\mathrm{inter}\}$.

\paragraph{Intra-regional and Inter-regional Trips}
The trips are classified into four categories based on origins and destinations: intra-regional trips within Region 1 ($\Theta^1_\mathrm{intra}$) and within Region 2 ($\Theta^2_\mathrm{intra}$), and inter-regional trips from Region 1 ($\Theta^1_\mathrm{inter}$) and from Region 2 ($\Theta^2_\mathrm{inter}$). 
The categorization of trips can be expressed as follows.
Given a request~$r_m=\tup{o_m,d_m,\alpha_m,\theta_m}$, one has:
\begin{align}
    (o_m \in \mathcal{V}_i \wedge d_m \in \mathcal{V}_i) \iff \theta_m = \Theta^i_\mathrm{intra}, \nonumber \\
    \forall i \in \mathcal{I}, m \in \mathcal{M} \nonumber,\\
    (o_m \in \mathcal{V}_i \wedge d_m \in \mathcal{V}_j \wedge i \neq j) \iff \theta_m = \Theta^i_\mathrm{inter}, \nonumber \\ 
    \forall i,j \in \mathcal{I}, m \in \mathcal{M} \nonumber.
\end{align}
The travel demand is generated using a discrete uniform distribution, denoted as $\alpha_m \sim \mathcal{U}(r^l,r^u)$, where the parameters of the lower $(r^u)$ and upper bounds $(r^l)$ can vary by trip type.  
In addition, to incorporate the impacts of population growth, we assume that travel demand will increase by a factor of $\tau$ annually.

\paragraph{Elastic Demand}
Elastic demand indicates that the choices of travelers are responsive to the quality of service provided, i.e., a higher performance of the mobility system will attract more users.
For each travel request, there are two preselected routes: a train-prioritized route and an alternative-mode-based route (the alternative route). 
The former prioritizes the use of train services, with alternative modes being used only when trains are not available. 
In contrast, the alternative route uses other transportation modes.
The edge sets for these routes are denoted as $\mathcal{E}^R_m$ and $\mathcal{E}^A_m$, respectively. 
Travelers choose between these options by evaluating the travel costs of both routes ($u^R_m$ and $u^A_m$), which are determined by summing the mobility service price and the monetary value of time:
\begin{align}
u^A_m
= &\sum_{e \in \mathcal{E}^A_m} (\gamma_\mathrm{vot} \frac{l_e}{v_A} + \gamma_\mathrm{A} l_e ), \quad \forall m \in \mathcal{M} \label{eq:uA}, \\
u^R_m
=&\sum_{e \in \mathcal{E}^R_m} l_e x_e(\frac{\gamma_\mathrm{vot}}{v_R} + \gamma_\mathrm{t}), \nonumber\\
+&(1-x_e) \sum\limits_{a \in \mathcal{E}_R} \mathds{1}_{\mathcal{E}^a_e} l_a(\frac{\gamma_\mathrm{vot}}{v_A} + \gamma_\mathrm{A}), 
\quad \forall m \in \mathcal{M},\label{eq:uR} 
\end{align} 
where $x_e$ is the service availability on edge $e$, $\gamma_\mathrm{vot}$ is the value of time, $\gamma_\mathrm{a}$ and $\gamma_\mathrm{t}$ are the distance-based price for using alternative modes and train service, respectively, $l_e$ is the travel distance on edge $e$, and $v_a$ and $v_t$ are the average speed of using alternative modes and the speed of trains, and $\mathds{1}_{\mathcal{E}^a_e}$ denotes whether the edge $e$ is part of the alternative-mode edges that travelers will take if the train service on edge $a$ is not available. 
%
For the request $r_m$, a proportion $p_m \in [0, 1]$ of trips will choose the train-prioritized route, which can be determined by: 
\begin{align}
&p_{m} = \frac{e^{-u^R_m}}{e^{-u^A_m} + e^{-u^{R}_m}}, \quad  \forall m \in \mathcal{M},\label{eq:p_m}
\end{align}
where  $u^{R}_m$ and $u^A_m$ denote the travel cost of the train-prioritized route and the alternative route, respectively. 



\paragraph{Interactive Network Design Problem}
Due to the existence of intercity trip requests, the performance of any subnetwork is influenced by the layout of all subnetworks, i.e., the decision of all regional authorities.
Regional authorities are assumed to be rational and aim to maximize their payoff. 
It is assumed that regional authorities make such decisions \emph{simultaneously}, without prior knowledge of the choices made by others.
In this work, the notion of Nash Equilibrium represents the convergence to a network where no regional authority can improve their objective at the specific stage by unilaterally changing their actions. 

 \paragraph{The Condition to Cooperate} 
Authorities may negotiate and reach possible agreements in their decision-making process.
Given that regional authorities are selfish agents, we further assume that each authority will evaluate the agreements based on their individual goals. 

\begin{definition}[Equilibrium of the interactive network design process]
\label{def:ne}
A strategy profile $\tup{H_i,H_{-i}}$ is a Nash Equilibrium of the interactive network design process if, for all regional authorities $i \in I$, $f_i(H_i,H_{-i}) \geq f_i(\hat{H}_i,H_{-i}) \ \forall$ $\hat{H}_i \in \mathcal{H}_i$.
\end{definition}

This indicates that no regional authority can improve their payoff by unilaterally changing their strategy from $H^t_i$ to any other strategy $\hat{H}^t_i$, given that the strategies of other authorities $H_{-i}$ remain unchanged. 
 
\begin{definition} [Feasible agreement] 
\label{def:feasible}
A feasible agreement is defined as the negotiated result that all the regional authorities agree on, satisfying the following condition:
\begin{align}
    f_i(H^\mathrm{co}) \geq f_i(H^\mathrm{ne}), \nonumber
\end{align}
where $H^\mathrm{co}$ represents the strategy in the agreement, and $H^\mathrm{ne}$ is the Nash Equilibrium strategy profile for the interactive network design process. 
The cooperation can only be formed if every authority can have higher payoff, compared to the payoff obtained at the Nash equilibrium in a non-cooperative game.
\end{definition} 


\subsection{Interactive Network Design Framework} \label{sec:coop}
In this section we introduce the model structure and variables, present the mathematical model for interactive network design in detail, along with the objective functions and constraints characterizing each regional stakeholder.

\subsubsection{Model structure and Variables}
The interactive network design game consists of two design stages integrating a non-cooperative game and a cooperative one.
The first stage is non-cooperative (detailed in \cref{sec:stage1}), where regional authorities independently develop strategic plans with individual goals without any negotiation.
During this phase, each player aims to choose the best strategy for themselves. 
Given the designed network in the first stage, in the second stage (detailed in \cref{sec:stage2}), players explore opportunities to enlarge the collective benefit and then seek ways to distribute it (i.e., ``making the pie larger'' and then ``dividing the pie fairly''). 
We assume that regional authorities will cooperate only when their benefit is higher than expected.

Given that network design is a long-term decision-making process typically spanning several years, we first define a planning horizon $T$, indexed by $t$.
We assume that the regional authories have the full information on the total number of travel requests and the annual travel demand growth rate $\tau$.
The decision variables are given by $H^t_i= \tup{x^t_e, s^{t}_e}_{e \in \mathcal{E}^i_{R}}\in \{0,1\} \times S_\mathrm{fre}$, where $x^t_e = 1 $ denotes the action of constructing a service connection on the edge $e$ during year $t$, and the integer variable $s^{t}_e \in S_\mathrm{fre}=\{0,1,...,s_\mathrm{max}\}$ indicates the action of upgrading the service frequency on edge $e$ during year $t$. 

The network layout variables for each year are given by $\mathcal{X}^t_i=\tup{X_e^t,S_e^t}_{e \in \mathcal{E}^i_R}\in \{0,1\} \times S_\mathrm{fre}$, where $X_e^t$ represents the connectivity of the edges during year $t$, and $S_e^t$ denotes the service frequency at the edges.
The layout of each year builds on the layout of previous years such that $\mathcal{X}^t_i = H_i^t + \mathcal{X}^{t-1}_i, t \in \mathcal{T}=\{1,2,3...,T\}$.  
The initial network configuration is denoted by $\mathcal{X}^0$, and serves as the foundation for the first-year network design ($t=1$).
The relation (\ref{eq:nd_sx}) ensures that no service frequency can be assigned to edge $e$ unless the train service becomes available ($X^{t}_e=1$), with $\Omega$ representing a large positive value.
We assumed that the service capacity is linearly related with the service frequency, shown in \cref{eq:nd_cs}:
\begin{align}
    &X_e^t= X_e^{t-1}+x_e^{t-1}\quad \forall t \in \mathcal{T}\label{updatex},\\
    &S_e^t= S_e^{t-1}+s_e^{t-1} \quad \forall t \in \mathcal{T} \label{updates},\\
    & S^{t}_e \leq X^{t}_e\Omega \quad \forall t \in \mathcal{T}, \label{eq:nd_sx}\\
    & c^t_e=\kappa S^t \quad \forall t \in \mathcal{T}.  \label{eq:nd_cs}
\end{align}
The variable $Y^t=(y^t_e)_{e \in \mathcal{E}}$ denotes the served flow on edges.
The flows on the railway edges $\mathcal{E}_R$ are determined by the elastic railway travel demand and restricted by the edge capacity $c_e$. The flows on the edges of the alternative layer $\mathcal{E}_A$ depend on both the demand from travelers selecting alternative routes and the unserved demand from the railway network.
Specifically, $y_e (\mathcal{X}^t)$ behaves as follows:
\begin{align}
    \begin{cases}
    \min\left\{\sum\limits_{m \in \mathcal{M}}  \mathds{1}_{\mathcal{E}^R_m(e)} \alpha_m p_m, c_e \right\}, & \text{if } e \in \mathcal{E}_R,\\
    \sum\limits_{m \in \mathcal{M}} \mathds{1}_{\mathcal{E}^A_m(e)} \alpha_m (1-p_m) + 
    \sum\limits_{a \in \mathcal{E}_R} \mathds{1}_{\mathcal{E}^a_e} \Delta_a, & \text{otherwise},
    \end{cases}
    \label{eq:y}
\end{align}
where $p_m$ is the proportion of travelers with request $m$ who are willing to choose the train-prioritized routes.
The boolean indicator functions $\mathds{1}_{\mathcal{E}^R_m(e)}$ and $\mathds{1}_{\mathcal{E}^A_m(e)}$ indicate whether edge $e$ is part of the edge set of the train-prioritized route and the alternative route, respectively. 
The amount of unserved travelers due to capacity constraints on the edge $a \in  \mathcal{E}_R$ is defined as:
\begin{align}
\Delta_a:=\max\left\{0,\sum\limits_{m \in \mathcal{M}} \alpha_m p_m  \mathds{1}_{\mathcal{E}^R_m({a})}-c_{a}\right\}. \nonumber
\end{align}
\subsubsection{Performance metrics} 
The performance of a transportation network can be evaluated from multiple perspectives. We assume that the regional authorities will consider the environmental, social, and economic impacts of various policies. Specifically, to assess network performance, regional authority $i$ calculates the CO\textsubscript{2} emissions, total travel costs, and profitability generated within its own region, denoted by $J^{e}_i$, $J^{c}_i$, and $J^{p}_i$ respectively. 
Such performance metrics depend not only on the network of their region but also on the network of the other one. 
The overall network layout is represented as $\mathcal{X}^t=(\mathcal{X}_i^t,\mathcal{X}_{-i}^t)$:
\begin{align}
    &J^{e}_i(\mathcal{X}^t)= 
        \sum_{e \in \mathcal{E}^i_R} \gamma_{m}^R l_e y^t_e
        + \sum_{e \in \mathcal{E}^i_A} \gamma_{m}^A l_e y^t_e,\\
    &J^{c}_i(\mathcal{X}^t)=  
        \sum_{e \in \mathcal{E}^i_R} l_e y^t_e( \frac{\gamma_\mathrm{vot}}{v_R}+\gamma_R)
         +\sum_{e \in \mathcal{E}^i_A} l_e y^t_e (\frac{\gamma_\mathrm{vot}}{v_A}+\gamma_A),\\
    &J^p_i(\mathcal{X}^t, H_i^t) = 
        \sum_{e \in \mathcal{E}^i_R}  \gamma_R l_e y^t_e 
        - \sum_{e \in \mathcal{E}^i_R} (c^{b} l_e x^t_e + c^{k} l_e s^{t}_e), \label{eq:Jp}
\end{align}
where the system emission $J^e_i$ accounts for both the train service and alternative services, positively related with the passenger amount and distance traveled.
The parameters $\gamma_{m}^R$ and $\gamma_{m}^A$ denote the CO\textsubscript{2} emission unit for train and alternative services, respectively. 
The total travel cost $J^{c}_i$ is the travel cost generated from all requests $\mathcal{R}$ within the region $i$.
In \cref{eq:Jp}, the profitability $J^p_i$ of the local network designer is the gap between the revenue from the train service and the construction cost. 
The construction cost includes both the base construction costs and the costs associated with upgrading the service frequency. The parameters $c^{b}$ and $c^{k}$ are the unit costs for line construction and frequency enhancement, respectively.

In \cref{sec:stage1} and \cref{sec:stage2}, we develop a interactive network design approach that models interactions and explores strategies that benefit both parties. 

\subsubsection{Design Stage 1 - Non-Cooperative Network Design} \label{sec:stage1}
The decision variables for the regional authority $i \in I =\{1,2\}$ are $H^{1t}_i= (x^t_e, s^{t}_e)_{e \in \mathcal{E}_{R}^i} \in \{0,1\} \times S_\mathrm{fre}$.
The regional authority $i$ optimizes the network design to maximize the objective (\ref{eq:nd_obj}), which is negatively associated with the system emission, and total travel cost and positively associated with profitability:
\begin{subequations} \label{problem:noncoop}
\begin{align}
    \max_{H^t_{i}} \quad & - \omega_0 J^e_i(\mathcal{X}_{s1}^t) - \omega_1 J^{c}_i(\mathcal{X}_{s1}^t) + \omega_2 J^p_i(\mathcal{X}_{s1}^t, H_i^{1t}) \label{eq:nd_obj}\\
    \text{s.t.}
    & \sum_{e \in \mathcal{E}^i_R} c^{b} l_e x^t_e +c^{k} l_e s^{t}_e \leq (1-\beta_i) B^\mathrm{t}_i
    \label{eq:nd_budget}, \\
    & \text{Eq. }(\ref{updatex})-(\ref{eq:y}) \nonumber,
\end{align}
\end{subequations}
where $\omega_0,\omega_1, \omega_2>0$ are weights.
$B^t_i$ indicates the budget of regional authority $i$ at the design stage $t$. 
Constraints in \cref{eq:nd_budget} ensure that the stage cost will not exceed the planned budget of operator $i$ in the year $t$, where $\beta_i$ denotes the budget rate for the cooperation stage. 
The designed network layout at Stage 1 is denoted as $\mathcal{X}_{s1}^t$.

\begin{remark}
The optimization problem (\ref{problem:noncoop}) is a mixed integer nonlinear programming problem, with binary decision variables $(x^t_e)_{e \in \mathcal{E}^R_i}$ and integer decision variables $(s^t_e)_{e \in \mathcal{E}^R_i}$. 
All constraints are linearly related to the decision variables. For each design period $t$ and regional authority $i$, the optimization problem has $\mathcal{O}\left(E_i\right)$ decision variables and $\mathcal{O}\left(E_i+M_i\right)$ constraints, where $E_i$ denotes the number of rail edges in the region $i$. $M_i:= \sum_{r_m \in \mathcal{R}} \alpha_m \mathds{1}\{o_m \in \mathcal{V}_i \vee d_m \in \mathcal{V}_i \}$ is the number of trips with the origin or the destination in the region $i$.
 \end{remark}

%
\paragraph*{Discussion on User-level Models for Network Design}
For the incorporation of user responses to both route and mode choices in the regional network design model, we propose an alternative user-level optimization model (see \cref{opt:ue}) to replace \cref{eq:y}. 
In line with UE traffic assignment, we formulate the following optimization problem for user-level modeling within the regional network design problem:
\begin{subequations} \label{opt:ue}
\begin{align}
    \min_{y_e \in [0,c_e]} &\sum_{e \in \mathcal{E}}\int_{0}^{y} g_e(y_e,x_e) \,dy,\\
    &\sum_{k \in \mathcal{K}} f_m^k=\alpha_{m},\\
    &f_{m}^{k} \geq 0,\\
    &y_e=\sum_{m \in \mathcal{M}} \sum_{k \in \mathcal{K}} f_{m}^{k}  \mathds{1}_{\mathcal{E}^k_m(e)},\\
    &y_e \leq c_e, \forall e \in \mathcal{E}_R, \label{ue:cap}\\
    & c_e<x_e \Omega, \forall e \in \mathcal{E}_R,
\end{align}
\end{subequations}
where $g_e (y_e,x_e)$ represents the edge travel cost in the multimodal transportation network. 
The constraints in \cref{ue:cap} ensure that the flow on the railway edge $e$ will not exceed the capacity $c_e$. 
The Bureau of Public Roads (BPR) function can be used to estimate the travel time of road traffic \cite{united1964traffic}. 
\begin{align}
    g_e (y_e,x_e) =
    \begin{cases}
    \gamma_{vot} t_{e_0}  (1+ a (\frac{y_e}{c_e})^{b})+l_e \gamma_A
    , & \text{if } e \in \mathcal{E}_A,\\
    x_el_e(\frac{\gamma_{vot}}{v_R}+\gamma_R)
    , &\text{otherwise}.
    \end{cases}
    \label{eq:g}
\end{align}
Then, the multi-regional network design problem can be structured as an Equilibrium Problem with Equilibrium Constraints (EPEC), considering the interaction between users at the lower level, and between network designers at the upper level.
In this work, we choose to simplify the user's decision-making model and use \cref{eq:y}.

\subsubsection{Design Stage 2 - Cooperative Network Design} \label{sec:stage2}
We develop two cooperative mechanisms for the cooperative network design: a co-investment mechanism and a payoff-sharing mechanism. 
The co-investment mechanism facilitates shared funding of rail services, and the payoff-sharing mechanism ensures that the generated benefits are fairly redistributed among authorities. 
\paragraph{Optimization Problem for Co-investment Mechanism}

The decision variables are $H^{2t}= (x^t_e, s^{t}_e)_{e \in \mathcal{E}_{R}} \in \{0,1\} \times S_\mathrm{fre}$, covering the entire network rather than a specific region.
Let $\mathcal{X}_{s2}^t$ be the network layout resulting from the co-investment mechanism, with $\mathcal{X}_{s1}^t \subseteq \mathcal{X}_{s2}^t$, indicating that the design in Stage 2 builds on the network in Stage 1. 
The optimization model (\ref{problem:coop_coinvest}) is presented to maximize the collective objectives of both regions, as follows:
\begin{subequations} \label{problem:coop_coinvest}
\begin{align}
    \max_{H^{t}} \quad & \sum_{i \in I} - \omega_0 J_i^e(\mathcal{X}_{s2}^t) 
    - \omega_1 J_i^c(\mathcal{X}_{s2}^t)\\
    &
    + \omega_2 J_i^p(\mathcal{X}_{s2}^t, H_i^{2t}) - F_i^{1t} \label{eq:co_obj}, \\
    \text{s.t.}\quad
    & \sum_{e \in \mathcal{E}_R} c^{b} l_e x^t_e +c^{k} l_e s^{t}_e \leq \sum_{i \in I} \beta_i B^\mathrm{t}_i \label{eq:co_budget}, \\
    & \text{Eq. }(\ref{updatex})-(\ref{eq:y}). \nonumber
\end{align}
\end{subequations}
The constraints in \cref{eq:co_budget} ensure that the co-investment does not exceed the allocated budget. 
$F_i^{1t}$ denotes the maximum objective obtained in Stage 1.

We use the \textit{co-investment ratio (CIR)} to represent the proportion of the total design budget allocated to co-investment
(${\sum_{i \in \mathcal{I}, t \in \mathcal{T}} \beta_i^t B_i^t}/{B}$), 
and \textit{return on co-investment (ROC)}  is used to represent the overall profitability of the co-investment mechanism ($\Delta F^\mathrm{co}/{\sum_{i \in \mathcal{I}, t \in \mathcal{T}} \beta_i^t B_i^t}$), where $\Delta F^\mathrm{co}$ is the additional payoff from the two-stage network design.
\begin{remark}
The optimization problem in \cref{problem:coop_coinvest} is a mixed integer nonlinear programming problem. The decision variables involve the entire rail network $(x^t_e)_{e \in \mathcal{E}^R}$, $(s^t_e)_{e \in \mathcal{E}^R}$. 
All constraints are linearly related to the decision variables. For each design period $t$ , the optimization problem has $\mathcal{O}\left(E\right)$ decision variables and $\mathcal{O}\left(E+M\right)$ constraints.
\end{remark}

\paragraph{Optimization Problem for Payoff-sharing Mechanism}
Payoff-sharing Mechanism is designed to distribute the payoffs generated through the co-investment mechanism. Let $\mathcal{Q} \subseteq \mathbb{R}^{I}$ be a closed and convex set of feasible payoff allocations, and let $d \subseteq \mathbb{R}^{I}$ be the disagreement point which are minimum payoffs of each agent if no agreement is reached. 

\begin{definition}
\label{def:nbs}
Nash Bargaining Solution (NBS), denoted by $v^*= \mathcal{N}(\mathcal{Q},d)=(v^*_i)_{i \in \mathcal{I}} \in \mathcal{Q}$, satisfies the following axioms:
\begin{enumerate}
    \item Pareto Optimality: $\forall v \in \mathcal{Q} \setminus v^*,$  if  $ \exists i \in \mathcal{I}, v_i > v_i^*  \Rightarrow \exists j \in \mathcal{I}, v_j < v_j^* $;
    \item Symmetry: If $\exists i,j \in \mathcal{I}, i \neq j,  d_i=d_j,  \Rightarrow v^*_i=v^*_j$;
    \item  Independence of Irrelevant alternatives: if $v^* \in \mathcal{Q}^{'}\subset \mathcal{Q} \Rightarrow \mathcal{N}(\mathcal{Q},d) = \mathcal{N}(\mathcal{Q}^{'},d)$;
    \item Independent of Linear Transformation: if  $\phi: \mathbb{R}^{I} \rightarrow  \mathbb{R}^{I} , \forall i \in \mathcal{I}, \phi_i(v_i)=a_iv_i+b_i, a_i >0 $ \\ $\Rightarrow \mathcal{N}(\phi(\mathcal{Q}),\phi(d))=\phi(\mathcal{N}(\mathcal{Q},d))$.
\end{enumerate}
\end{definition}
In this work, we apply the NBS concept for the payoff mechanism modeling. 
The disagreement payoffs are derived from the outcomes of non-cooperative games $d=(F^{t}_i)_{i \in \mathcal{I}} \in  \mathbb{R}^+$ indicating that if the co-investment and payoff-sharing can not lead to additional benefits for both parties, the negotiations fail and they behave non-cooperatively.

Based on \cref{def:nbs}, NBS for the pay-off mechanism ensures that the resulting payoff allocation must satisfy Pareto efficiency, where no reallocation can increase the payoff of one local authority without decreasing the payoff of the other. The agreement fully exploits the potential payoffs available to both authorities.
The Symmetry Axiom ensures that regional authorities are not specially labeled. 
The independence of Irrelevant Alternatives principle implies that the payoff-sharing decision must remain unaffected by changes in the set of feasible solutions (NBS remains within the feasible set) and is independent of other feasible alternatives.
In addition, the bargaining result must be invariant under affine transformations of the payoff.

Payoff-sharing mechanism can be formulated as the following optimization problem (\ref{problem:coop_payoffshare}): 
\begin{subequations} \label{problem:coop_payoffshare}
\begin{align}
     \max_{q_i^t \in \mathcal{Q}} \quad &  \prod_{i \in I} v^t_i-F_i^{t} \label{ps:obj}\\
     \text{s.t. }
     &v^t_i > F^{t}_i, \quad \forall i \in I ,\label{eq:v-disagreement}\\
     &v^t_i=F^{1t}_{i}+q_i, \quad \forall i \in I,\\
     &\sum_{i \in I} q_i^t=\sum_{i \in I}F^{2t}_{i}. \label{eq:v-pool}
\end{align}
\end{subequations}
where $q_i^t \in \mathcal{Q}^t$ denotes the distributed payoff to authority $i$ through the payoff-sharing mechanism, and $v^t_i$ represents the payoff of regional authority $i$ from the interactive network design in year $t$.
$F^{1t}_{i}$ represents the payoff for local authority $i$ in the non-cooperative stage (maximum objective value of Model (\ref{problem:noncoop})), $F^{2t}_{i}$ denotes the payoff resulting from the co-investment mechanism (maximum objective value of Model (\ref{problem:coop_coinvest})). $F_i^{t}$ is the payoff in the scenario where no mechanism is applied, with regional authorities only focusing on their respective regions (maximum objective value of Model (\ref{problem:noncoop}) with $\beta_i=0$).
The constraints in \cref{eq:v-disagreement} ensure that the interactive network design can result in a higher payoffs for both regional authorities.
The constraints in \cref{eq:v-pool} restrict the sharable value to the amount generated by the co-investment mechanism.

\begin{lemma} [Existence and uniqueness of optimal payoff allocations]
\label{coro:payoff}
The payoff mechanism can produce an unique optimal solution for payoff allocations by the optimization problem, only if 
\begin{align}
    \sum_{i \in \mathcal{I}} F^{1t}_{i} +F^{2t} >\sum_{i \in \mathcal{I}} F^{t}_i \label{eq:corollary1}.
\end{align}
\end{lemma}
This inequality indicates that the payoff mechanism will provide a solution only when the co-investment yields a higher total payoff than in the case of complete non-cooperation.

\begin{proof}
The condition \(\sum_{i \in \mathcal{I}} F^{1t}_{i} +F^{2t} >\sum_{i \in \mathcal{I}} F^{t}_i\) ensures that the set \(Q^t\) is a non-empty, convex, and compact subset of \(\mathbb{R}^I\). 
Because $F^{1t}_{i}$ is a constant, the function \(f:Q^t \rightarrow V^t\) is a concave and upper-bounded function defined on \(Q^t\). 
The objective function (Equation (\ref{ps:obj})) is continuous and strictly quasi-concave with respect to the decision variable \(q_i^t\), under the constraint that \(\sum_{i \in \mathcal{I}} q_i^t = F^{2t}\).
Therefore, the optimal solution exists for the optimization problem (\ref{problem:coop_payoffshare}).
Since \(f\) is injective, the optimal solution is unique.
\end{proof}



\section{Numerical Experiments} \label{sec:exp}
\begin{figure}[tb]
    \centering
    \begin{tikzpicture}[scale=1.3]
    \definecolor{steelblueline}{RGB}{167,200,227}
    \tikzstyle{node_style1}=[circle, draw, fill={rgb,255:red,255;green,200;blue,150}, minimum size=5mm, inner sep=0pt, font=\tiny]
    \tikzstyle{node_style2}=[circle, draw, dashed, fill=steelblueline, minimum size=5mm, inner sep=0pt, font=\tiny]
    \def \radius {10cm}
    \node[node_style2] (13) at (0, 0) {13};
    \node[node_style2] (24) at (1, 0) {24};
    \node[node_style2] (21) at (2, 0) {21};
    \node[node_style2] (20) at (3, 0) {20};
    \node[node_style2] (23) at (1, 0.7) {23};
    \node[node_style2] (22) at (2, 0.7) {22};
    \node[node_style2] (14) at (1, 1.4) {14};
    \node[node_style2] (15) at (2, 1.4) {15};
    \node[node_style2] (19) at (3, 1.4) {19};
    \node[node_style2] (17) at (3, 2.1) {17};
    \node[node_style2] (12) at (0, 2.8) {12};
    \node[node_style1] (11) at (1, 2.8) {11};
    \node[node_style1] (10) at (2, 2.8) {10};
    \node[node_style2] (16) at (3, 2.8) {16};
    \node[node_style2] (18) at (4, 2.8) {18};
    \node[node_style1] (7) at (4, 3.5) {7};
    \node[node_style1] (8) at (3, 3.5) {8};
    \node[node_style1] (9) at (2, 3.5) {9};
    \node[node_style1] (6) at (3, 4.2) {6};
    \node[node_style1] (5) at (2, 4.2) {5};
    \node[node_style1] (4) at (1, 4.2) {4};
    \node[node_style1] (3) at (0, 4.2) {3};
    \node[node_style1] (2) at (3, 4.9) {2};
    \node[node_style1] (1) at (0, 4.9) {1};

    \tikzstyle{line_style}=[->,  line width=0.5pt, font=\fontsize{5}{6}\selectfont]
    \draw[line_style] (2) -- (1) node[midway, above]{3};
    \draw[line_style] (1.south east) -- (2.south west) node[midway, below] {1};
    \draw[line_style] (4) -- (3) node[midway, above]{8};
    \draw[line_style] (5) -- (4) node[midway, above]{11};
    \draw[line_style] (6) -- (5) node[midway, above]{15};
    \draw[line_style] (7) -- (8) node[midway, above]{17};
    \draw[line_style] (8) -- (9) node[midway, above]{21};
    \draw[line_style] (18) -- (16) node[midway, above]{55};
    \draw[line_style] (16) -- (10) node[midway, above]{48};
    \draw[line_style] (10) -- (11) node[midway, above]{27};
    \draw[line_style] (11) -- (12) node[midway, above]{33};
    \draw[line_style] (19) -- (15) node[midway, above]{57};
    \draw[line_style] (15) -- (14) node[midway, above]{41};
    \draw[line_style] (22) -- (23) node[midway, above]{72};
    \draw[line_style] (20) -- (21) node[midway, above]{62};
    \draw[line_style] (21) -- (24) node[midway, above]{66};
    \draw[line_style] (24) -- (13) node[midway, above]{74};
    \draw[line_style] (1.south west) -- (3.north west) node[midway, left]{2};
    \draw[line_style] (12.south west) -- (13.north west) node[midway, left]{37};
    \draw[line_style] (4.south west) -- (11.north west) node[midway, left]{10};
    \draw[line_style] (11.south west) -- (14.north west) node[midway, left]{34};
    \draw[line_style] (14.south west) -- (23.north west) node[midway, left]{42};
    \draw[line_style] (23.south west) -- (24.north west) node[midway, left]{73};
    \draw[line_style] (5.south west) -- (9.north west) node[midway, left]{13};
    \draw[line_style] (9.south west) -- (10.north west) node[midway, left]{25};
    \draw[line_style] (10.south west) -- (15.north west) node[midway, left]{28};
    \draw[line_style] (15.south west) -- (22.north west) node[midway, left]{46};
    \draw[line_style] (22.south west) -- (21.north west) node[midway, left]{69};
    \draw[line_style] (2.south west) -- (6.north west) node[midway, left]{4};
    \draw[line_style] (6.south west) -- (8.north west) node[midway, left]{16};
    \draw[line_style] (8.south west) -- (16.north west) node[midway, left]{22};
    \draw[line_style] (16.south west) -- (17.north west) node[midway, left]{49};
    \draw[line_style] (17.south west) -- (19.north west) node[midway, left]{53};
    \draw[line_style] (19.south west) -- (20.north west) node[midway, left]{59};
    \draw[line_style] (3.south west) -- (12.north west) node[midway, left]{7};

    \draw[line_style] (1.south east) -- (2.south west) node[midway, below] {1};
    \draw[line_style] (3.south east) -- (4.south west) node[midway, below]{6};
    \draw[line_style] (4.south east) -- (5.south west) node[midway, below]{9};

    \draw[line_style] (5.south east) -- (6.south west) node[midway, below]{12};
    \draw[line_style] (8.south east) -- (7.south west) node[midway, below]{20};
    \draw[line_style] (9.south east) -- (8.south west) node[midway, below]{24};
    \draw[line_style] (16.south east) -- (18.south west) node[midway, below]{50};
    \draw[line_style] (10.south east) -- (16.south west) node[pos=0.45, below]{29};
    \draw[line_style] (11.south east) -- (10.south west) node[midway, below]{32};
    \draw[line_style] (12.south east) -- (11.south west) node[midway, below]{36};
    \draw[line_style] (15.south east) -- (19.south west) node[midway, below]{45};
    \draw[line_style] (14.south east) -- (15.south west) node[midway, below]{44};
    \draw[line_style] (23.south east) -- (22.south west) node[midway, below]{72};
    \draw[line_style] (21.south east) -- (20.south west) node[midway, below]{64};
    \draw[line_style] (24.south east) -- (21.south west) node[midway, below]{75};
    \draw[line_style] (23) -- (14) node[midway, right]{74};
    \draw[line_style] (3) -- (1) node[midway, right]{5};
    \draw[line_style] (12) -- (3) node[midway, right]{35};
    \draw[line_style] (13) -- (12) node[midway, right]{38};
    \draw[line_style] (11) -- (4) node[midway, right]{31};
    \draw[line_style] (14) -- (11) node[midway, right]{40};
    \draw[line_style] (13.south east) -- (24.south west) node[midway, below]{39};
    \draw[line_style] (24) -- (23) node[midway, right]{76};
    \draw[line_style] (9) -- (5) node[midway, right]{23};
    \draw[line_style] (10) -- (9) node[midway, right]{26};
    \draw[line_style] (15) -- (10) node[midway, right]{43};
    \draw[line_style] (22) -- (15) node[midway, right]{67};
    \draw[line_style] (21) -- (22) node[midway, right]{65};
    \draw[line_style] (6) -- (2) node[midway, right]{14};
    \draw[line_style] (8) -- (6) node[midway, right]{19};
    \draw[line_style] (16) -- (8) node[midway, right]{47};;
    \draw[line_style] (17) -- (16) node[midway, right]{52};
    \draw[line_style] (19) -- (17) node[midway, right]{58};
    \draw[line_style] (20) -- (19) node[midway, right]{61};
    \draw[line_style] (18.south) -- (20.north east) node[midway, left]{56};
    \draw[line_style] (20.east) -- (18.south east) node[midway, right]{60};
    \draw[line_style] (7.south west) -- (18.north west) node[midway, left]{18};
    \draw[line_style] (18) -- (7) node[midway, right]{54};
    \draw[line_style] (20) -- (22.east) node[midway,above]{63};
    \draw[line_style] (22.south east) -- (20.west) node[pos=0.35, below ]{68};
    \draw[line_style] (17) -- (10.south east)  node[pos=0.35, above]{51};
    \draw[line_style] (10.south) -- (17.west) node[midway, below]{30};
    \node[draw=black, fill={gray!8!white}, minimum width=2.8cm, minimum height=1.5cm] at (4.58, 4.75) {};

    \node[node_style1] at  (3.8, 5) {};
    \node[draw=none, minimum size=3mm,fill=none, anchor=west] at (4, 5) {\scriptsize Nodes of Region 1};
    \node[node_style2] at (3.8, 4.5) {};
    \node[draw=none, minimum size=3mm,fill=none, anchor=west] at (4, 4.5) {\scriptsize Nodes of Region 2};

\end{tikzpicture}
    \caption{Sioux Falls network, subdivided between Region 1 and 2.}
    \label{fig:sioux}
\end{figure}
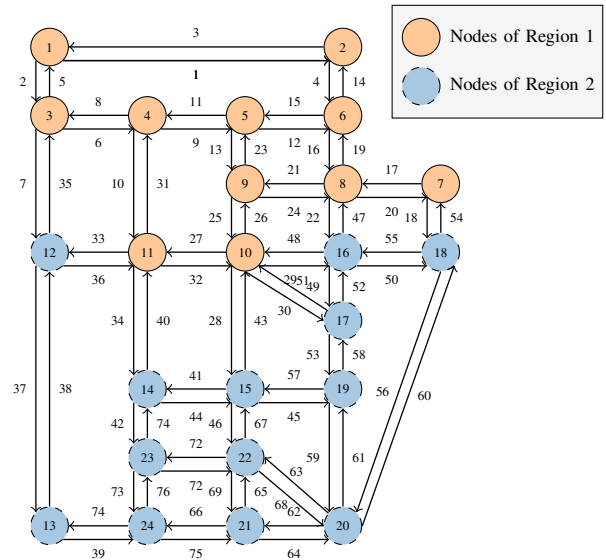

To demonstrate the proposed framework, we conduct numerical experiments on the well-known Sioux Falls network in the USA, as shown in \cref{fig:sioux}. The network consists of 24 nodes and 76 edges. 
We divide it into two regions, with nodes 1-11 assigned to Region 1 and nodes 12-24 to Region 2.
The parameters for the model, transportation modes, and travelers are listed in \cref{tab:Parameters} in Appendix \ref{appendix:para}. 

First, we investigate the case in which regions are homogeneous in terms of construction resources and travel demand distributions. 
We simulate and optimize the network design using the proposed framework
The baseline solution is with ($\beta^t_i=0$), indicating that no cooperative mechanism is adopted.
\begin{figure*}[h!]
    \centering
    \includegraphics[width=0.8\linewidth]{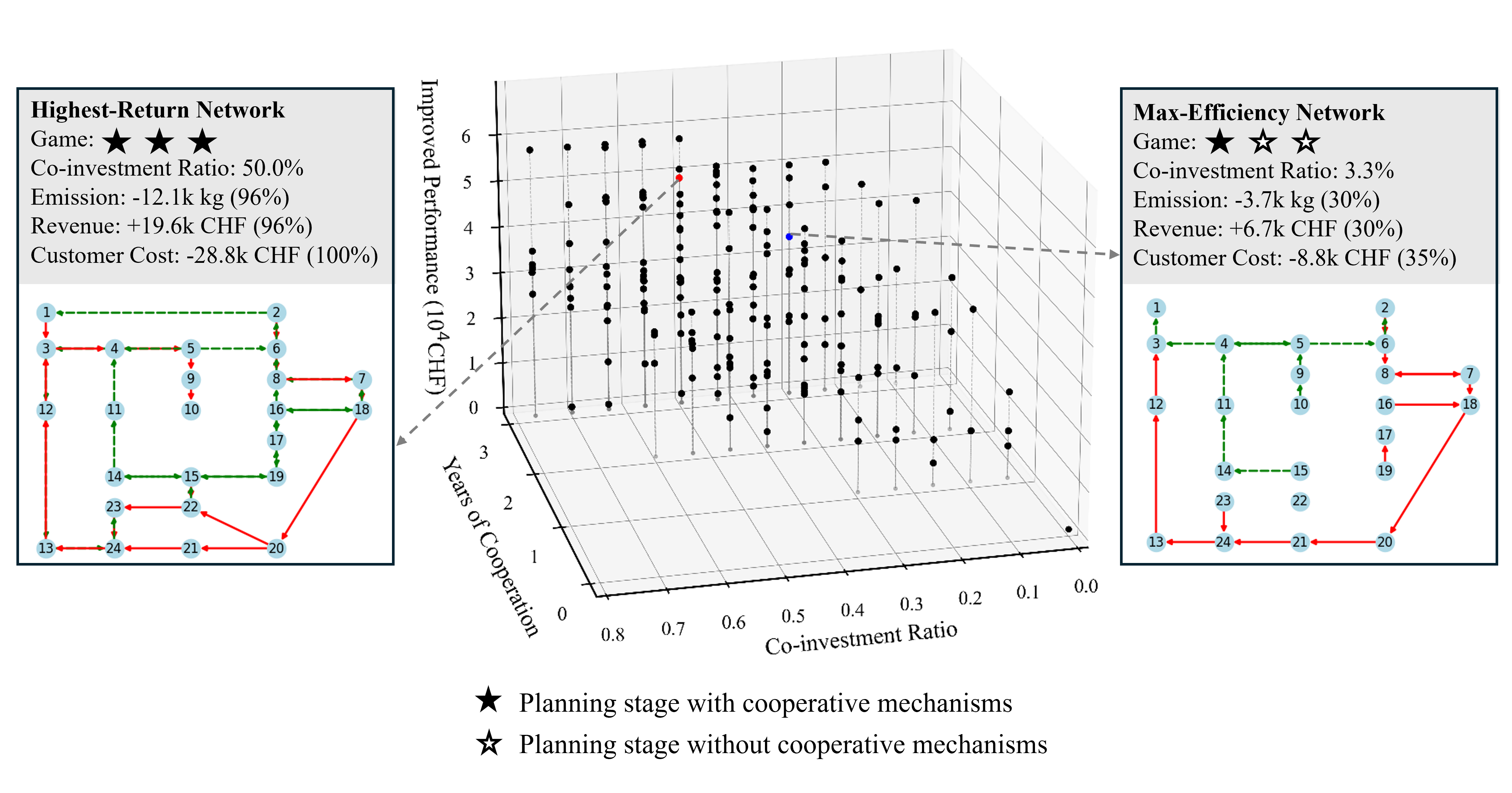}
    \caption{Equilibrium solutions of interactive network design.}
    \label{fig:overview}
\end{figure*}
\cref{fig:overview} shows the equilibrium solutions for the railway network after three years of design, where each solution involves varying budget allocations for co-investment over years. 
The axis ``years of cooperation'' indicates the number of years the regional authorities chose to adopt the proposed mechanisms. The improvement in the objective thanks to the cooperative mechanism is measured in CHF per day. 
To assess how well the system approaches the system-optimal solution (centralized design), we measure the percentage of improvement in environmental, social, and economic outcomes achieved by the equilibrium solutions.
Two solutions are shown in detail for illustration.
The filled star indicates the decision to co-invest during the network design phrase. The presented networks highlight the differences between the specific cases and the baseline network: edges in red indicate more resources allocated, while green edges represent less constructed edges.
The red dot represents the Highest-Return Network, which indicates the equilibrium network with the highest improvement in the objective. This solution requires continuous co-investment in each design year, with the co-investment budget accounting for 50\% of the total budget. As a result, emissions can be reduced by 12.1 ton/day, revenue increases by 19.6k CHF/day, and customer costs are reduced by 28.8k CHF/day. With the proposed mechanism, allocating 50\% of the budget for co-investment can achieve outcomes that are very close to the optimal system results, reaching 96\%, 96\%, and 100\% across the three dimensions.
The blue dot represents the solution where the design strategy yields the most investment-efficient outcome. By co-investing only 3.3\% of the total budget in the initial design year, the system achieves a 3.7 ton/day reduction in emissions, 8.8k CHF/day in travel cost savings, and an additional revenue of 6.7k CHF/day.
\begin{figure}[tb]
    \centering
    \input{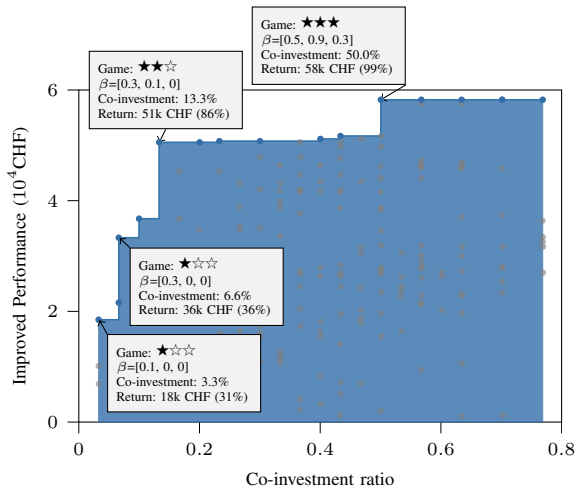}
    \caption{Co-investment ratio and improved performance.}
    \label{fig:co_ratio}
\end{figure}
It can be observed from \cref{fig:co_ratio} that, with the same co-investment ratio, the performance varies depending on the year of investment and the distributed funding. 
For instance, by allocating 30\% of the budget in the first year and 10\% in the second year for co-investment, the resulting network can approach the system-optimal solution by 86\% with an additional return of 51k CHF/day, exceeding other budget distribution strategies with the same total investment.

We can also examine the effects of the interactive network design framework in scenarios involving heterogeneous regional authorities. With the total budget and travel demand held constant, we adjusted the budget and the intra-city travel in Region 2 (parameters shown in \cref{tab:hete} in Appendix \ref{appendix:para}). The figure presents the range of return on co-investment resulting from the interactive network design.
Overall, for the network design with heterogeneity regions, incorporating cooperation within non-cooperation results in higher RoC for the system. The scenario yielding the highest RoC occurs when Region 2 has a larger budget and fewer intra-city trips compared to Region 1.
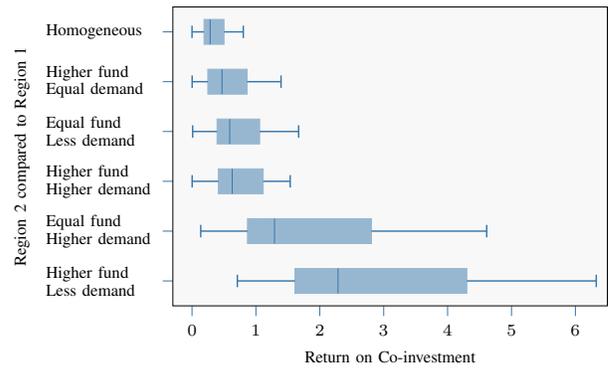
\begin{figure}[tb]
    \centering
\begin{tikzpicture}[scale=0.9]

\definecolor{darkgray176}{RGB}{176,176,176}
\definecolor{steelblueline}{RGB}{55,118,171}
\definecolor{steelblueface}{RGB}{55,118,171}
\begin{axis}[
    width=8cm,  
    height=6cm, 
    tick align=outside,
    tick pos=left,
    x grid style={darkgray176},
    xlabel={Return on Co-investment},
    xmin=-0.3, xmax=6.5,
    xtick style={color=steelblueline},
    y grid style={darkgray176},
    ylabel={Region 2 compared to Region 1},
    ymin=0.5, ymax=6.5,
    ytick style={color=steelblueline},
    ytick={1,2,3,4,5,6},
    yticklabels={
      {\parbox{1.6cm}{Higher fund \\ Less demand}},
      {\parbox{1.6cm}{Equal fund \\ Higher demand}},
      {\parbox{1.6cm}{Higher fund \\ Higher demand}},
      {\parbox{1.6cm}{Equal fund\\ Less demand}},
      {\parbox{1.6cm}{Higher fund\\Equal demand}},
      \parbox{1.6cm}{Homogeneous}
    },
    font=\scriptsize,
    axis background/.style={fill=lightgray,opacity=0.1},
    ]

\path [draw=steelblueline, fill=steelblueline,opacity=0.5]
(axis cs:1.61103373015873,0.75)
--(axis cs:1.61103373015873,1.25)
--(axis cs:4.30059583333333,1.25)
--(axis cs:4.30059583333333,0.75)
--(axis cs:1.61103373015873,0.75)
--cycle;
\addplot [semithick, steelblueline]
table {%
1.61103373015873 1
0.709388888888889 1
};
\addplot [semithick, steelblueline]
table {%
4.30059583333333 1
6.32665 1
};
\addplot [semithick, steelblueline]
table {%
0.709388888888889 0.875
0.709388888888889 1.125
};
\addplot [semithick, steelblueline]
table {%
6.32665 0.875
6.32665 1.125
};
\path [draw=steelblueline, fill=steelblueline,opacity=0.5]
(axis cs:0.868380357142857,1.75)
--(axis cs:0.868380357142857,2.25)
--(axis cs:2.80814583333333,2.25)
--(axis cs:2.80814583333333,1.75)
--(axis cs:0.868380357142857,1.75)
--cycle;
\addplot [semithick, steelblueline]
table {%
0.868380357142857 2
0.136366666666667 2
};
\addplot [semithick, steelblueline]
table {%
2.80814583333333 2
4.61175 2
};
\addplot [semithick, steelblueline]
table {%
0.136366666666667 1.875
0.136366666666667 2.125
};
\addplot [semithick, steelblueline]
table {%
4.61175 1.875
4.61175 2.125
};
\path [draw=steelblueline, fill=steelblueline,opacity=0.5]
(axis cs:0.412950793650794,2.75)
--(axis cs:0.412950793650794,3.25)
--(axis cs:1.11391666666667,3.25)
--(axis cs:1.11391666666667,2.75)
--(axis cs:0.412950793650794,2.75)
--cycle;
\addplot [semithick, steelblueline]
table {%
0.412950793650794 3
0.00216 3
};
\addplot [semithick, steelblueline]
table {%
1.11391666666667 3
1.53673333333333 3
};
\addplot [semithick, steelblueline]
table {%
0.00216 2.875
0.00216 3.125
};
\addplot [semithick, steelblueline]
table {%
1.53673333333333 2.875
1.53673333333333 3.125
};
\path [draw=steelblueline, fill=steelblueline,opacity=0.5]
(axis cs:0.391859523809524,3.75)
--(axis cs:0.391859523809524,4.25)
--(axis cs:1.058425,4.25)
--(axis cs:1.058425,3.75)
--(axis cs:0.391859523809524,3.75)
--cycle;
\addplot [semithick, steelblueline]
table {%
0.391859523809524 4
0.0102 4
};
\addplot [semithick, steelblueline]
table {%
1.058425 4
1.6684 4
};
\addplot [semithick, steelblueline]
table {%
0.0102 3.875
0.0102 4.125
};
\addplot [semithick, steelblueline]
table {%
1.6684 3.875
1.6684 4.125
};
\path [draw=steelblueline, fill=steelblueline,opacity=0.5]
(axis cs:0.2470625,4.75)
--(axis cs:0.2470625,5.25)
--(axis cs:0.86129,5.25)
--(axis cs:0.86129,4.75)
--(axis cs:0.2470625,4.75)
--cycle;
\addplot [semithick, steelblueline]
table {%
0.2470625 5
0.00214285714285714 5
};
\addplot [semithick, steelblueline]
table {%
0.86129 5
1.39331666666667 5
};
\addplot [semithick, steelblueline]
table {%
0.00214285714285714 4.875
0.00214285714285714 5.125
};
\addplot [semithick, steelblueline]
table {%
1.39331666666667 4.875
1.39331666666667 5.125
};
\path [draw=steelblueline, fill=steelblueline,opacity=0.5]
(axis cs:0.187461111111111,5.75)
--(axis cs:0.187461111111111,6.25)
--(axis cs:0.501033333333333,6.25)
--(axis cs:0.501033333333333,5.75)
--(axis cs:0.187461111111111,5.75)
--cycle;
\addplot [semithick, steelblueline]
table {%
0.187461111111111 6
0.000142857142857143 6
};
\addplot [semithick, steelblueline]
table {%
0.501033333333333 6
0.80295 6
};
\addplot [semithick, steelblueline]
table {%
0.000142857142857143 5.875
0.000142857142857143 6.125
};
\addplot [semithick, steelblueline]
table {%
0.80295 5.875
0.80295 6.125
};
\addplot [steelblueline]
table {%
2.285245 0.75
2.285245 1.25
};
\addplot [steelblueline]
table {%
1.29114428571429 1.75
1.29114428571429 2.25
};
\addplot [steelblueline]
table {%
0.62961 2.75
0.62961 3.25
};
\addplot [steelblueline]
table {%
0.590025555555556 3.75
0.590025555555556 4.25
};
\addplot [steelblueline]
table {%
0.469296428571429 4.75
0.469296428571429 5.25
};
\addplot [steelblueline]
table {%
0.28367 5.75
0.28367 6.25
};
\end{axis}

\end{tikzpicture}
    \caption{The impact of interactive network design for heterogeneous regions.}
    \label{fig:profitAB}
\end{figure}
\section{Conclusion}  \label{sec:conclusion}
In this work, we proposed a game-theoretic framework for the interactive network design problem. 
In the network design framework, the design process is divided into two stages: in the first stage, regional authorities focus their own networks and individual objectives; in the second, the co-investment and payoff-sharing mechanisms are developed. 
A case study on the Sioux Falls network demonstrates the framework's properties. 
The results show that interactive network design can improve rail network performance from environmental, social and economic perspectives. 

For future research, the co-investment mechanism can be enhanced by incorporating decisions on the co-investment ratio instead of imposing a fixed upper bound.
In addition, the temporal dimension can be explored. In this context, questions of interest include how one decision on cooperation can impact subsequent decisions. 
From the point of view of case studies, we want to investigate larger, inter-city networks.
\bibliographystyle{IEEEtran}
\bibliography{ref}

\begin{appendices}

\section{Parameters for network design model and experiment scenarios}
\label{appendix:para}
\cref{tab:Parameters} presents model parameters, and \cref{tab:hete} shows the parameters used in the heterogeneous-region scenarios.
\begin{table}[!h]
    \centering
    \caption{Model Parameters}
    \scalebox{0.8}{
    \begin{tabular}{lllll}
    \hline
        Parameters& Description & Value  & Unit & Reference \\ \hline
        \multicolumn{5}{l}{\textbf{Network design}} \\
         $T$ & Planning horizon & 3 &  year & - \\ 
        $B_i^t$ & Budget & $1 \times 10^5$ &  CHF/day & - \\ 
        $\beta^t_i$ & \multicolumn{4}{l}{Regional co-invest ratio \{0, 0.1, 0.3, 0.5, 0.7, 0.9\}}  \\
        $c^{b}$ & Base cost & 574 &  CHF/day/km & \cite{basecost} \\ 
        $c^{k}$ & Capacity cost & 31.4 & CHF/day/km & \cite{bosch2018cost} \\ 
        $s_\mathrm{max}$ & Maximum frequency & 15 & veh/h& ~\\ 
        $a,b$ & BPR function & 4, 0.15 & -& \cite{united1964traffic}\\
        $\Omega$ & Large number& $1 \times 10^5$ & -& -\\
        \multicolumn{5}{l}{\textbf{Travel demand}} \\
        $\tau$ & Demand growth rate & 1.5 &\% & \cite{growthrate}\\
        $\gamma_\mathrm{vot}$ & Value of time & 30  & CHF/h & \cite{schmid2021value} \\ 
        \multicolumn{5}{l}{\textbf{Mobility service}} \\
        \textbf{Train} & ~ & ~ & ~ & ~ \\ 
        $\gamma_{c}$ & Service fee & 0.25  & CHF/km/pax & \cite{trainfare} \\ 
        $\gamma_{m}^R$ & Emission & 0.019 & kg/km/pax & \cite{em} \\ 
        $\gamma_{R}$ & Speed & 150 & km/h & \cite{trainspeed} \\ 
        $\kappa$ & Capacity & 500  & seat/veh & \cite{cap} \\
        \multicolumn{5}{l}{\textbf{Alternative mode}} \\
        $\gamma_{t}$ & Service fee & 1.65 & CHF/km/pax & \cite{taxifare} \\ 
        $\gamma_{m}^A$ & Emission & 0.148 & kg/km/pax & \cite{em} \\ 
         $\gamma_{R}$ & Speed & 100 & km/h & \cite{taxispeed} \\ \hline
    \end{tabular}
    }
    \label{tab:Parameters}
\end{table}

\begin{table}[!h]
    \centering
    \caption{Scenario Parameters for co-investment and travel demand}
    \scalebox{0.8}{
    \begin{tabular}{lcc}
    \hline
     Scenarios & $B_1:B_2$ & $|R(\Theta_1^{intra})|:|R(\Theta_2^{intra})|$   \\ \hline
     Homogeneous & 1:1 & 1:1  \\ 
     Higher fund, Equal pop& 3:2 & 1:1\\
     Equal fund, Less pop & 1:1 & 2:3\\
     Higher fund, Higher pop& 3:2 & 3:2\\
     Equal fund, Higher pop& 1:1 & 3:2\\
     High fund,  Less pop& 3:2 & 2:3\\
    \hline
    \end{tabular}
    }
    \label{tab:hete}
\end{table}
\end{appendices}

\end{document}